\documentstyle[psfig]{cupconf}

\ifoldfss
\else
  \ifnfssone
    \newmathalphabet{\mathit}
      \addtoversion{normal}{\mathit}{cmr}{m}{it}
      \addtoversion{bold}{\mathit}{cmr}{bx}{it}
    \newmathalphabet{\mathcal}
      \addtoversion{normal}{\mathcal}{cmsy}{m}{n}
    \else
    \ifnfsstwo
    \fi
  \fi
\fi

\def\kms{$\rm km\;s^{-1}$}
\def\hexnumber#1{\ifcase#1 0\or1\or2\or3\or4\or5\or6\or7\or8\or9\or
 A\or B\or C\or D\or E\or F\fi }

\makeatletter
\ifx\CUP@mtlplain@loaded\undefined
\else
\fi
\makeatother
 \makeatletter
 \ifx\CUP@mtlplain@loaded\undefined
   \font\tenbmi=cmmib10 at 10pt
   \font\sevenbmi=cmmib10 at 7pt
   \font\fivebmi=cmmib10 at 5pt

   \newfam\bmifam
   \textfont\bmifam=\tenbmi
   \scriptfont\bmifam=\sevenbmi
   \scriptscriptfont\bmifam=\fivebmi
   
 \fi
 \makeatother
\ifnfsstwo

\fi
\ifnfssone

\fi
\ifoldfss

\fi

\mathchardef\varLambda="0103

\makeatletter
\ifx\CUP@mtlplain@loaded\undefined
\else
\fi
\makeatother
\makeatletter
\ifx\CUP@mtlplain@loaded\undefined
  \font\tenbms=cmbsy10
  \font\sevenbms=cmbsy10 at 7pt
  \font\fivebms=cmbsy10 at 5pt
  \newfam\bmsfam
  \textfont\bmsfam=\tenbms
  \scriptfont\bmsfam=\sevenbms
  \scriptscriptfont\bmsfam=\fivebms

  \edef\bsy@{\hexnumber\bmsfam}
  \mathchardef\bnabla="0\bsy@72
\fi
\makeatother
\title[NGC~4698 and NGC~4672]{The bulge-disk orthogonal decoupling in 
galaxies: NGC~4698 and NGC~4672}

\author[F. Bertola {\it et al.\/}]%
{F.\ns B\ls E\ls R\ls T\ls O\ls L\ls A$^1$,\ns
 E.\ns M.\ns C\ls O\ls R\ls S\ls I\ls N\ls I$^1$,\ns\\
 M.\ns C\ls A\ls P\ls P\ls E\ls L\ls L\ls A\ls R\ls I$^1$\ns
 J.\ns C.\ns V\ls E\ls G\ls A\ns B\ls E\ls L\ls T\ls R\ls \'A\ls N$^2$,\ns\\
 A.\ns P\ls I\ls Z\ls Z\ls E\ls L\ls L\ls A$^3$,\ns
 M.\ns S\ls A\ls R\ls Z\ls I$^1$,\ns
 \and \ns 
 J.\ns G.\ns F\ls U\ls N\ls E\ls S\ns S.\ns J.$^1$} 

\affiliation{$^1$Dipartimento di Astronomia, Universit\`a di Padova,
Vicolo dell'Osservatorio 5, I-35122 Padova, Italy\\[\affilskip]
$^2$Telescopio Nazionale Galileo, Osservatorio Astronomico
di Padova, Vicolo dell'Osservatorio 5, I-35122 Padova, Italy\\[\affilskip]
$^3$ European Southern Observatory, Alonso de Cordova 3107,
Casilla 19001, Santiago 10, Chile}

\begin{document}
\ifnfssone
\else
  \ifnfsstwo
  \else
    \ifoldfss
      \let\mathcal\cal
      \let\mathrm\rm
      \let\mathsf\sf
    \fi
  \fi
\fi

\maketitle

\begin{abstract}
We report the case of the geometrical and kinematical decoupling
between the bulge and the disk of the Sa galaxy NGC~4698.  The
$R-$band isophotal map of this spiral shows that the bulge structure
is elongated perpendicularly to the major axis of the disk.  At the
same time a central stellar velocity gradient is found along the major
axis of the bulge. We also present the Sa NGC~4672 as good candidate
of a spiral hosting a bulge and a disk orthogonally decoupled with
respect to one other.  This decoupling of the two fundamental
components of a visible galaxy suggests that the disk could represent
a second event in the history of early-type spirals.
\end{abstract}

\firstsection % if your document starts with a section,

\section{Introduction}

NGC~4698 is classified Sa by Sandage \& Tammann (1981) and Sab(s) by
de Vaucouleurs {\it et al.\/} (1991, RC3). Sandage \& Bedke (1994, CAG)
present NGC~4698 as an example of the early-to-intermediate Sa type
since it is characterized by a large central bulge and tightly wound
spiral arms. In addition to a remarkable geometrical decoupling
between the bulge and the disk whose apparent major axes appear
oriented in an orthogonal way at simple visual inspection of galaxy
plates (see Panels 78, 79 and 87 in CAG), a spectrum taken along the
minor axis of the disk shows the presence of a stellar velocity 
gradient which could be ascribed to the bulge. For this reason
NGC~4698 is a noteworthy case for the study of the formation processes
of disks in spirals.

\section{Results}
  
The $R-$band isophotal map of NGC~4698 (Fig.~1 left panel) shows the
geometrical decoupling ($\rm \Delta\,P.A.\,\simeq\,90^\circ$) between
the bulge and the disk of these spiral. The phenomenon is visible both
in the inner isophotes and in the outermost one, which is
characterized by two `bumps' protruding perpendicularly to the galaxy
major axis. The isophotes between $4''$ and $19''$ appear round in
the plot. However as soon as an exponential disk is subtracted, they
become elongated perpendicularly to the disk major axis.  In order to
disentangle between the light contribution of bulge and disk, we
decomposed the surface-brightness radial profiles extracted along
different axes of NGC~4698 as the sum of an $r^{1/4}$ bulge ($\mu_e =
19.6$ mag$\cdot$arcsec$^{-2}$; $r_e = 11.3''$; $q\,=\,1.1$) and an
exponential disk ($\mu_0 = 19.2$ mag$\cdot$arcsec$^{-2}$; $r_d =
32.2''$; $i=\,60^\circ$).  The bulge results the dominating component
inside $10''$ along the galaxy major axis and inside $14''$ along the
minor axis. The axial ratio of the bulge is found to be greater than
unity confirming its exceptional property to be elongated along the
disk minor axis.  Alternative non-parametric decompositions of the
NGC~4698 surface-brightness distribution (Moriondo {\it et al.\/} 1998) with a
disk profile flattening toward the center would produce lower
residuals when the modeled surface brightness is subtracted from the
observed one.  However our standard assumption of an $r^{1/4}$ bulge
and an exponential disk does not affect the photometric and
kinematical orthogonal decoupling between bulge and disk of NGC~4698
(Fig.~1).

\begin{figure}
\vspace{7cm}
\includegraphics{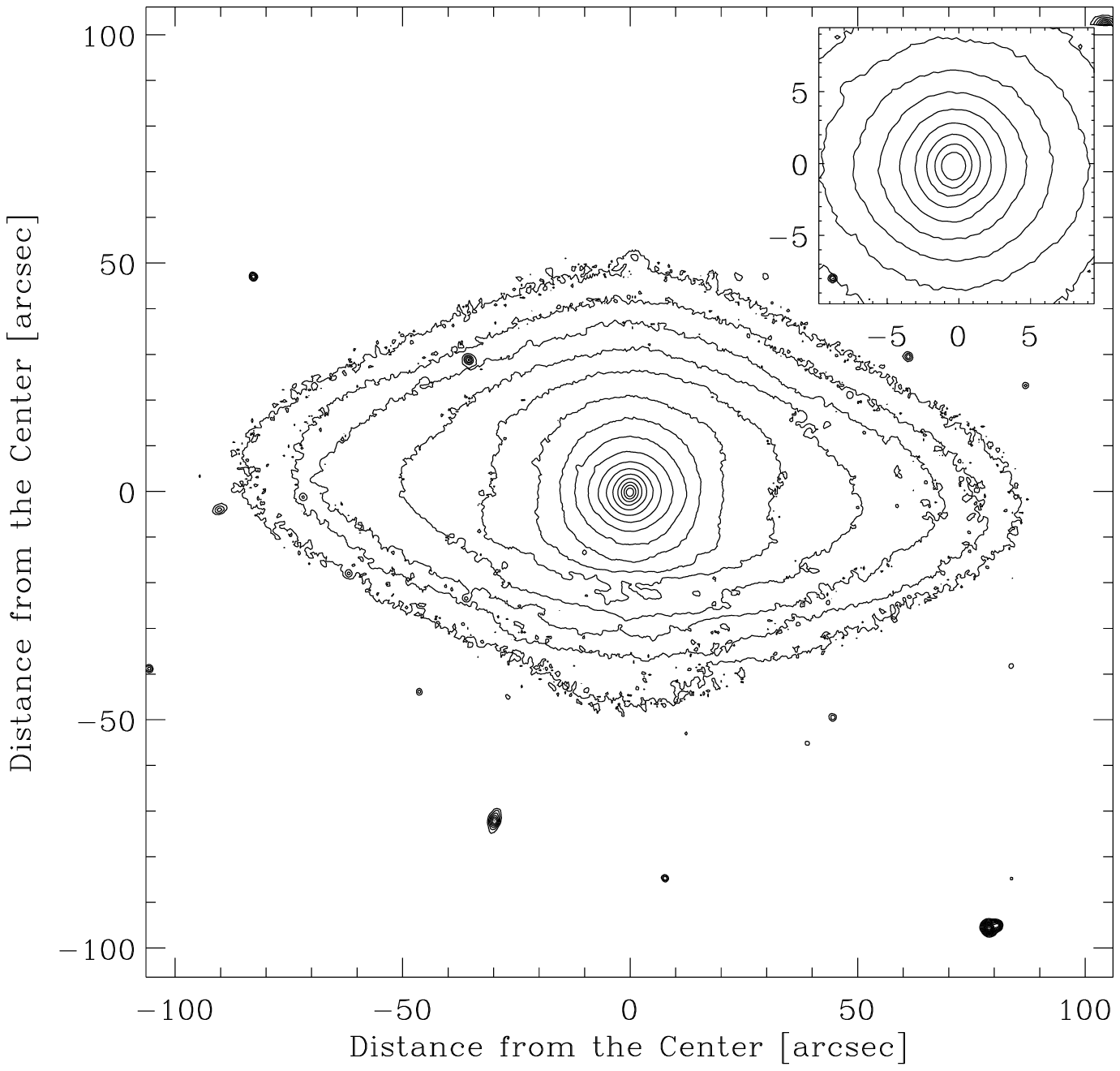}
\includegraphics{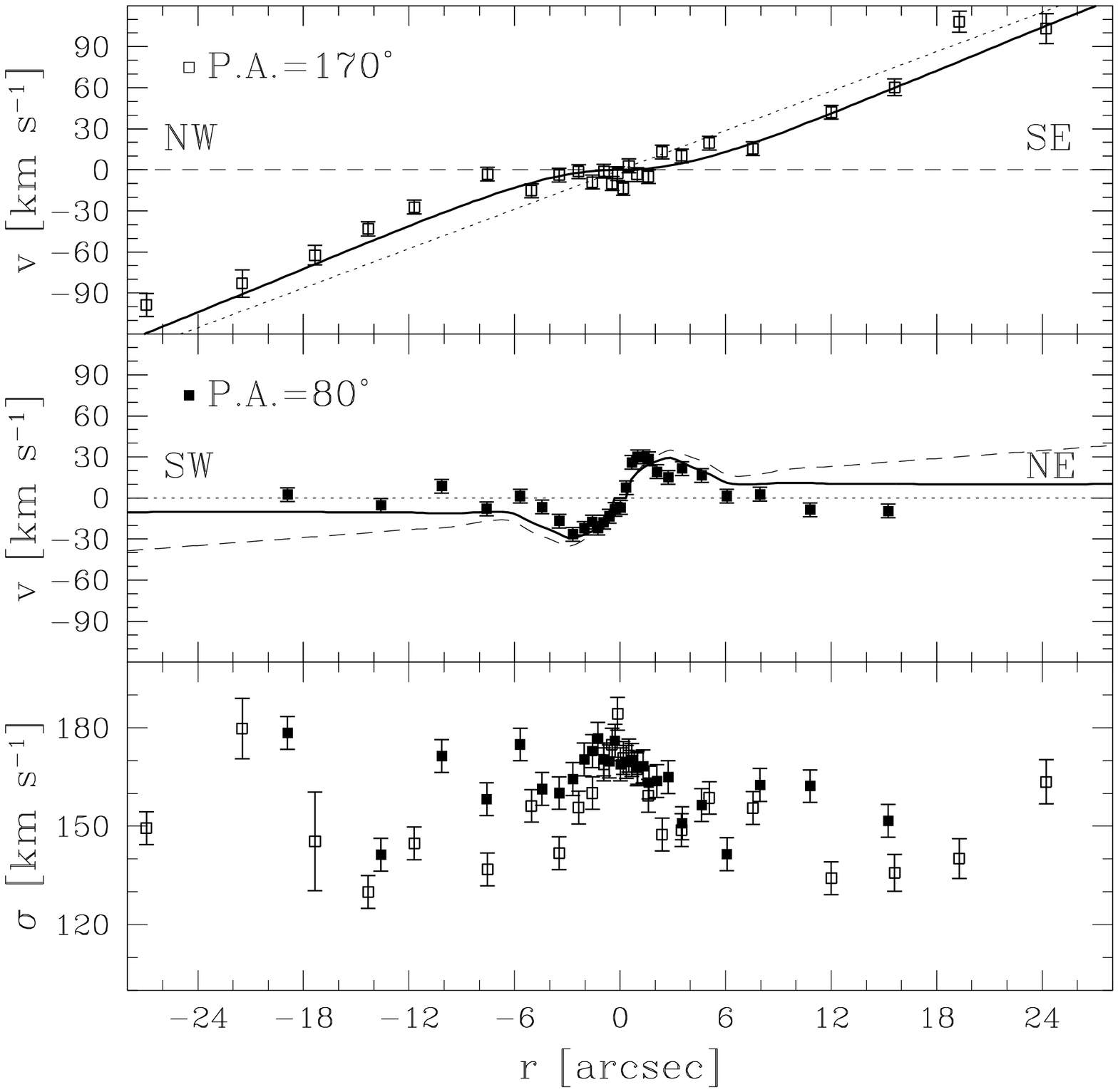}
\caption{{\it Left panel:\/} $R-$band isophotes of NGC~4698 given in
  steps of 0.4 mag$\cdot$arcsec$^{-2}$ with the outermost one
  corresponding to 21.8 mag$\cdot$arcsec$^{-2}$ and the central one to
  15.8 mag$\cdot$arcsec$^{-2}$. In the inset the isophotal map of the
  inner $10''$ is plotted. North is to right and east up.  
  {\it Right panel:\/} Observed stellar rotation velocity ($V_\odot =
  992 \pm 10$ \kms) and velocity dispersion as a function of radius
  along the major (open squares) and minor (filled squares) axes of
  NGC~4698.  The dashed and the dotted lines represent the velocity
  contribution of the bulge and disk components to the total velocity
  (thick continuous line) of our model}
\end{figure}

The major-axis stellar velocity curve is characterized by a central
plateau, indeed the stars have a zero rotation for
$|r|\,\leq\,8''$. At larger radii the observed stellar rotation
increases from zero to an approximately constant value of about 200
\kms\ for $|r|\,\geq\,50''$ up to the farthest observed radius at
about $80''$. These velocities are in agreement within the errors with
those measured by Corsini {\it et al.\/} (1999).  The stellar velocity
dispersion profile has been measured out to $30''$. It is peaked in
the center at the value of 185 \kms.  We measured the minor-axis
stellar kinematics out to about $20''$ on both sides of the galaxy. In
the nucleus the stellar velocity rotation increases to about 30 \kms\
at $|r|\,\simeq\,2''$, decreasing to zero further out.  The velocity
dispersion profile has a central maximum of 175 \kms\ in agreement
with the value measured on the spectrum along the major axis. The
stellar velocity curves and velocity-dispersion radial profiles (out
only to $28''$ for the spectrum along the major axis) are plotted in
Fig.~1 (right panel).
In order to demonstrate that the observed velocity curves along the
major and minor axis are consistent with the rotation of a bulge and a
disk with perpendicular angular momenta we modeled the observed
line-of-sight velocity distribution in the following way. Along the
major axis we assumed a velocity curve for the bulge constant at zero
velocity with a constant velocity dispersion at $\sigma_b\,=\,160$
\kms, while for the velocity curve of the disk we assumed the velocity
rising linearly to match the outer points of the plotted curve where
the light contribution of the bulge is negligible, with a constant
velocity dispersion at $\sigma_d\,=\,130$ \kms. The resulting velocity
curve is obtained by fitting with a Gaussian the sum of the two
velocity components corresponding to the bulge and the disk weighted
according to the photometric decomposition. The agreement with the
observed points is very good and the flat central part of the observed
velocity curve is well reproduced. A similar approach has been applied
to reproduce the velocity curve along the minor axis. A constant zero
velocity and a constant velocity dispersion $\sigma_d\,=\,130$ \kms\
have been assumed for the disk, as well as a constant velocity
dispersion $\sigma_b\,=\,160$ \kms\ for the bulge. The bulge rotation
has been maximized in such a way that the resulting velocity curve is
after folding within the scatter of the data. It reaches a maximum of
35 \kms\ in the inner $3''$ decreasing to a local minimum of 15 \kms\
at $|r|\,\simeq\,6''$ and then increasing afterwards.

\section{Discussion and conclusion}

The observed stellar kinematics can be interpreted as due to an
orthogonal kinematical decoupling between the bulge and disk
components. Assuming that the intrinsic shape of a bulge is generally
triaxial (Bertola, Vietri \& Zeilinger 1991), and that the plane of
the disk coincides with the plane of the bulge perpendicular either to
the major or to the minor axis, we deduce that the observed
configuration indicates that the major axis of the bulge is
perpendicular to the disk, given that the disk is seen not far from
edge on. The fact that the velocity field of the bulge is
characterized by a zero velocity along its apparent minor axis, as
indicated by the central plateau in the rotation curve along the disk
major axis, and by a gradient along its major axis suggests that the
rotation axis of the bulge lies on the plane of the disk.

Our photometric and spectroscopic data and the ensuing interpretation
of the orthogonal decoupling between the bulge and disk of NGC~4698
can be explained if the disk has formed in a distinct process occurred
in the history of the galaxy.  We suggest that the disk has formed at
a later stage due to acquisition of material by a pre-existing
triaxial spheroid on its principal plane perpendicular to the major
axis. An example of acquisition on the plane perpendicular to the
minor axis could be represented by NGC~7331 (Prada {\it et al.\/} 1996), where
the bulge has been found counterrotating with respect to the disk.  Up
to now NGC~4698 and NGC~7331 represent the only cases of kinematical
evidence that disk galaxies with prominent bulge could be started as
`undressed spheroid' and their disks accreted gradually over several
billion years, as suggested by Binney \& May (1986). Recently such
kind of processes have been considered within semi-analytical modeling
techniques for galaxy formation, where the disks accrete around bare
spheroids previously formed (e.g. Kauffmann 1996; Baugh, Cole \& Frenk
1996). In this framework polar-ring elliptical galaxies like NGC~5266
(Varnas {\it et al.\/} 1987) and ellipticals with dust lanes along the minor
axis (Bertola 1987) could be a transient stage towards the formation
of spiral systems like NGC~4698.

In order to verify if the acquisition phenomena giving rise to
NGC~4698 and NGC~7331 are general processes of galaxy formation we
need to know how these objects are unique or peculiar. To address this
question we began a photometric and spectroscopic survey of early-type
spirals even with a slight indication of geometrical orthogonal
decoupling between bulge and disk. The finding of galaxies with round
bulges suggests that the case of the orthogonal geometrical decoupling
between the bulge and disk of NGC~4698 could be a more general
phenomenon. In fact, even a bulge which appears round on a plate
becomes intrinsically elongated after the subtraction of an inclined
exponential disk. The further spectroscopic analysis of this kind of
objects with the slit of the spectrograph set along the major axis of
their bulges allow to determine if the geometrical decoupling is
associated with the kinematical decoupling, as in NGC~4698.
As first result of this survey we present the case of NGC~4672 (Sarzi
{\it et al.\/} 1999, in preparation). It is very probably a highly-inclined Sa
spiral with a prominent $r^{1/4}$ bulge sticking out from the plane of
the disk (e.g. Fig.~3i by Whitmore {\it et al.\/} 1990) rather than a
polar-ring galaxy.  The stellar velocity field (Fig.~3) is
characterized by the signature of a central zero-velocity plateau in
the rotation curve obtained along the major-axis ($\rm
P.A.\,=\,46^{\circ}$) and by a steep velocity gradient observed along
the minor axis ($\rm P.A.\,=\,134^{\circ}$) similar to those measured
in NGC~4698. In addition the major-axis gas rotation curve is typical
of a disk in differential rotation and not of a ring.  The close
structural and kinematical resemblance of NGC~4672 to NGC~4698 makes
the former a good candidate to be a new case of early-type spiral
characterized by a bulge-to-disk orthogonal decoupling.

\begin{figure}
\vspace{7cm}
\includegraphics{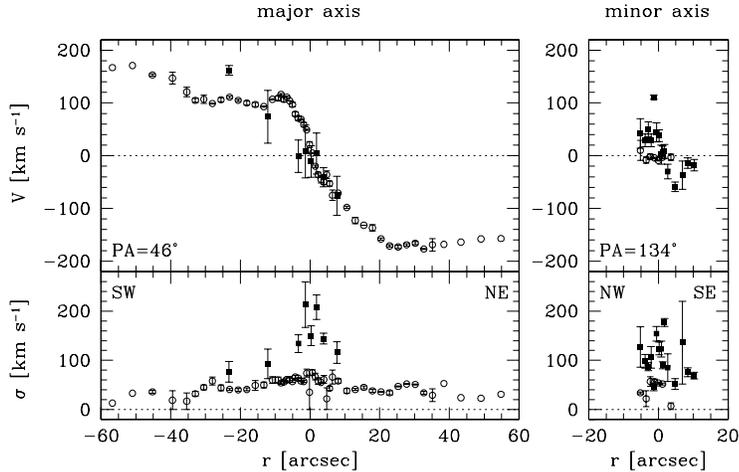}
\caption{The stellar (filled squares) and ionized gas (open circles)
  kinematics measured along the major ($\rm P.A.\,=\,46^{\circ}$) and minor
  axis ($\rm P.A.\,=\,134^{\circ}$) of NGC~4672. The systemic velocity is
  $V_\odot = 3275 \pm 20$ \kms}
\end{figure}

%\begin{acknowledgments}
%\end{acknowledgments}

\end{document}